# DESIGN OF MECHANISMS FOR ENSURING THE EXECUTION OF TASKS IN PROJECT PLANNING

*This paper reports an analysis of aspects of the project planning stage. The object of research is the decision-making processes that take place at this stage. This work considers the problem of building a hierarchy of tasks, their distribution among performers, taking into account restrictions on financial costs and duration of project implementation.*

*Verbal and mathematical models of the task of constructing a hierarchy of tasks and other tasks that take place at the stage of project planning were constructed.*

*Such indicators of the project implementation process efficiency were introduced as the time, cost, and cost-time efficiency. In order to be able to apply these criteria, the tasks of estimating the minimum value of the duration of the project and its minimum required cost were considered. Appropriate methods have been developed to solve them.*

*The developed iterative method for assessing the minimum duration of project implementation is based on taking into account the possibility of simultaneous execution of various tasks. The method of estimating the minimum cost of the project is to build and solve the problem of Boolean programming.*

*The values obtained as a result of solving these problems form an «ideal point», approaching which is enabled by the developed iterative method of constructing a hierarchy of tasks based on the method of sequential concessions. This method makes it possible to devise options for management decisions to obtain valid solutions to the problem. According to them, the decision maker can introduce a concession on the value of one or both components of the «ideal point» or change the input data to the task.*

*The models and methods built can be used when planning projects in education, science, production, etc.*

**Oksana Mulesa**
*Corresponding author*
Doctor of Technical Sciences, Professor*
E-mail: Oksana.mulesa@uzhnu.edu.ua

**Petro Horvat**
PhD, Associate Professor, Head of Department
Department of Computer Systems and Networks**

**Tamara Radivilova**
Doctor of Technical Sciences, Professor
V. V. Popovskyy Department of Infocommunication Engineering
Kharkiv National University of Radio Electronics
Nauky ave., 14, Kharkiv, Ukraine, 61166

**Volodymyr Sabadosh**
Senior Java Developer
Intellias Company
Kapushanska str., 177 A, Uzhhorod, Ukraine, 88000

**Oleksii Baranovskyi**
PhD, Senior Lecturer
Department of Computer Science
Blekinge Institute of Technology
Karlskrona, Blekinge, Sweden, 37179

**Sergii Duran**
Postgraduate Student*
*Department of Software Systems**
**Uzhhorod National University
Narodna sq., 3, Uzhhorod, Ukraine, 88000

## 1. Introduction

Project management is an important mechanism for ensuring the efficiency of companies, regardless of their type and field of activity [1]. Science, production, business in the modern world is project-oriented [2–4]. The key to the success of the creation and implementation of the project is the effective execution of its planning stage. At this stage, one usually analyzes possible options for the implementation of the project itself, evaluates the available resources, draws up a work plan [5, 6]. At the same time, there are problems associated with the need to comply with time and financial constraints, efficient use of labor resources, etc. Among the tasks that are solved in solving these problems are the problems of combinatorial optimization [7], single and multicriteria optimization problems [8, 9], decision-making problems [10, 11], etc.

The ultimate criterion for the effectiveness of the drawn-up project plan is to meet the requirements of the problem owner or decision maker (DM).

The study of tasks that arise at the stage of project planning, the development of new methods for solving them will increase the efficiency of relevant decision-making processes.

## 2. Literature review and problem statement

In general, the task of building a hierarchy of tasks when planning projects can be attributed to the problems of combinatorial optimization [7]. Studies into problems of this type are reported in several modern scientific papers [12, 13]. To solve them, depending on the nature of the input data, different classes of methods have been developed. These methods, with increasing dimensionality of input data and the number of constraints, have a sufficiently large computational complexity and do not always guarantee finding the optimal solution. The task of project planning in terms of the distribution of work between performers can be mathematically represented in the form of an appointment problem, as shown in [14, 15]. These studies present modern approaches to solving this problem. However, they do not take into account the possibility of consistent or simultaneous involvement of workers in different types of work. Also, an additional issue, when applying these methods, may be the problem of limited resources, which leads to the need for additional research. In [16], a gradient approach to solving a quadratic assignment problem is proposed. It makes it possible to take into account the priority of tasks and generates effective solutions in real time. However, similar to the previous approaches, it does not take into account the possibility of simultaneous involvement of performers in the implementation of different stages, as well as the issue of efficient use of resources.

Taking into account the need to ensure the efficient use of available resources, compliance with the project timeframe and other additional constraints that DM may provide, the project planning task can be represented as a multi-criteria optimization problem. One of the approaches to solving problems of multicriteria optimization is the convolution of all criteria into one [17–19]. This makes it possible to turn a multicriteria problem into a single-criteria problem and solve it using one of the known methods. The rules for constructing a convolution take into account the hierarchy of criteria by giving them weighting coefficients. This entails the need to rank a set of criteria and additionally calculate or specify their weights. To solve these problems, additional involvement of experts is possible, as shown in [20, 21]. This approach creates the need to solve additional problems, and also does not allow monitoring the value of each individual criterion.

In addition, to solve problems of multicriteria optimization, among others, evolutionary methods of multimodal optimization are applied [22, 23]. These methods generate a large number of optimal Pareto solutions. The choice of solution lies with the DM and can be a complex process. The task of choosing the optimal alternative from a set of given refers to decision-making problems. In [10], a method of decision-making in multi-stage decision-making processes is proposed. The method makes it possible, at the previous stages, to take into account the future consequences of decisions. This approach requires an assessment of all the consequences of adopting alternatives from a set of given ones, which is effective only with a small number of them.

For the case when the Pareto set has a sufficiently large power, it is proposed to apply dialog methods of multicriteria optimization, including the method of successive concessions [24, 25]. The use of such methods makes it possible to take into account the limitations provided by DM in a dialog mode, which significantly speeds up the decision-making process.

Therefore, it is advisable to design a solution tool that would allow the DM to take into account any number of performance criteria and, if necessary, make changes to the input data values in a dialog mode.

### 3. The aim and objectives of the study

The aim of this study is to design mechanisms to ensure the implementation of tasks when planning projects in compliance with time and financial constraints. This will increase the efficiency of management decisions, taking into account the priorities of the decision-maker.

To accomplish the aim, the following tasks have been set:
– to build a verbal and mathematical model and formalize the tasks that arise at the stage of project planning in the context of the distribution of performers between tasks;
– to devise methods for estimating the minimum values of the duration and cost of project implementation and based on them, develop an algorithm for an iterative method for building a hierarchy of tasks.

### 4. The study materials and methods

The object of our research is the decision-making processes that take place at the project planning stage. The problem of building a hierarchy of tasks and their distribution among performers with the imposition of additional restrictions on financial resources and duration of the project implementation stage is considered. In this case, a special case is in which the set of valid solutions is empty.

In the process of performing this study, methods of decision theory were used, which in a dialog mode allow the decision maker or the owner of the problem to achieve the desired values of the main parameters. In particular, the method of successive concessions was taken as a basis, according to which, if it is impossible to reach an ideal point, a compromise solution is worked out that degrades the values of some parameters to improve the values of other parameters of the system. This approach makes it possible to find solutions in cases where the initial set of valid solutions is empty.

In the course of the study, the following designations were introduced. Let $P$ be the project to be implemented and characterized by a tuple (1):

$$P =< A, C, S, W, R, T >, \qquad (1)$$

where $A$ – a set of tasks, the composition of the results of which ensures the implementation of the entire project, $A=\{A_1, A_2,..., A_N\}$; $C$ – the sum of costs for the implementation of the project $P$, $C \in R^+$ – positive real number; $S$ – ordered vector, the elements of which characterize the types of work to be performed by performers during project implementation, $S=(S_1, S_2,..., S_Q)$; $W$ – a set of potential executors in the project, $W=\{W_1, W_2,..., W_M\}$; $R$ – ordered vector, the elements of which determine the types of material and technical resources required for use during project implementation, $R=(R_1, R_2,..., R_K)$; $T$ – maximum allowable value of the duration of the project implementation stage.

Every element $A_i$, $i=\overline{1,N}$, of the set $A$ is represented as a tuple of the following form (2):

$$A_i =< IA_i, SA_i, RA_i, \Delta t_i, C_i >, \qquad (2)$$

where $IA_i$ – the set of task numbers that must be completed before the start of the task $A_i$, $IA_i \subset \{1, 2, ..., N\}$; $SA_i$ – ordered vector, the elements of which characterize the volume of individual work that must be performed during the task $A_i$, $SA_i=(s_{i1}, s_{i2},..., s_{iQ})$, where $s_{iq}$ is a real non-negative number equal to the amount of work $S_q$ in units of time, which must be done when performing task $A_i$, $q=\overline{1,Q}$; $RA_i$ is an ordered vector, the elements of which characterize the amount of resources in units of time that need to be involved in the course of the task $A_i$, $RA_i=(r_{i1}, r_{i2},..., r_{iK})$, where $r_{ik}$ is a non-negative integer equal to the volume of material and technical resources of the form $R_k$, which must be involved in the performance of task $A_i$, $k=\overline{1,K}$; $\Delta t_i$ – time required to complete task $A_i$; $C_i$ – the cost of implementing the task $A_i$.

For each performer $W_j \in W$, $j=\overline{1,M}$, we define a tuple (3):

$$<SW_j, CW_j>, \qquad (3)$$

where $SW_j=(w_{j1}, w_{j2},...,w_{jQ})$ is a Boolean vector, and $w_{jq}=1$ if the worker $W_j$ can perform work $S_q$ and $w_{jq}=0$ otherwise; $CW_j=(c_{j1}, c_{j2},..., c_{jQ})$ – vector, the components of which are real non-negative numbers and determine the cost of work performed by an employee per unit of time.

Set A can have different topologies. Some types of topologies are shown in Fig. 1; the arrow indicates the required sequence of tasks.

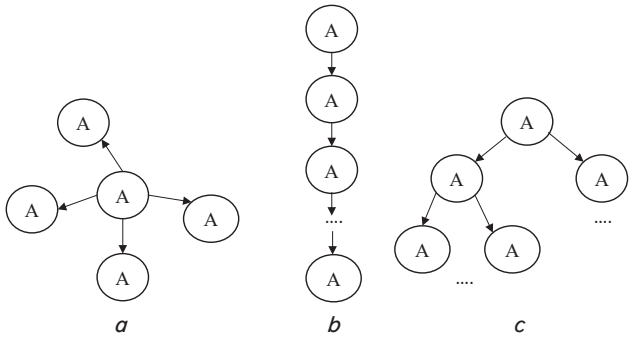

Fig. 1. Variants of topologies of set *A*:
*a* – «Star»; *b* – «Straight line»; *c* – «Tree»

The hierarchy of tasks, which will be built as a result of applying the developed methods, will become the basis of the project implementation process. Depending on the conditions of project implementation and the subjective characteristics of the problem owner or decision-maker, the following conditions may be the criteria for the effectiveness of the project implementation process.

Time efficiency consists in observing time constraints or minimizing the duration of all tasks involving specified human and logistical resources.

Cost efficiency is to comply with cost constraints or minimize the cost of performing all tasks.

Cost-time efficiency – compliance with restrictions on the cost and duration of the project.

## 5. Results of designing the mechanisms for ensuring the implementation of tasks in project planning

### 5. 1. Formalization of tasks that arise at the project planning stage

The tasks that arise at the planning stage of the project were formalized:

1. The task of building a hierarchy of tasks.

Under the hierarchy of tasks, we understand such an ordering of set *A*, in which each subsequent task can begin to be performed no earlier than all previous tasks began to be performed.

That is, in the simplest case, «build a hierarchy of tasks» means to specify the following ordering (4) for which conditions (5) hold:

$$A_{i_1}, A_{i_2},..., A_{i_N}, \qquad (4)$$

$$\bigcup_{v=1}^{N} A_{i_v} = A, \forall v \in \{2,3,...N\}\ IA_{i_v} \subseteq \{i_1, i_2,...,i_{v-1}\}. \qquad (5)$$

Obviously, $IA_{i_1} = \varnothing$.

The task of building a hierarchy of tasks is to specify such an ordering (4), following which, in the process of project implementation, the specified criteria for its effectiveness will be achieved.

2. The task of estimating the minimum value of the duration of the project is to find such a time period $\widetilde{T}$ that is enough to perform all tasks from set *A* with given human and logistical resources. It is clear that the maximum value of quantity $\widetilde{T}$ can be calculated from (6):

$$\widetilde{T}_{\max} = \sum_{i=1}^{N} \Delta t_i. \qquad (6)$$

However, in some cases it is possible to reduce the value of this quantity.

3. The task of estimating the minimum required cost of the project implementation process is to find such a minimum value of the cost $\widetilde{C}$, which is sufficient for the successful implementation of the project with specified human and material and technical resources.

### 5. 2. An iterative method for building a hierarchy of tasks based on the method of sequential assignments

At the first stage of studying the problem of constructing a hierarchy of tasks, the problem of estimating the duration range of the project implementation stage is solved. As already defined above, the maximum duration $\widetilde{T}$ is calculated from formula (6). To determine the lower boundary of the considered range, an iterative method for estimating the minimum value of the project duration has been developed.

The algorithm of the method is as follows:

Step 1. We fix the conditional start of the project $t_0=0$ and iteration number $iter=1$, and the list of already completed tasks $FA=\varnothing$.

Step 2. From the elements of set *A*, we construct the set $A^{(iter)}$ including those elements that can already begin to be executed in this iteration, that is,

$$A^{(iter)} = \bigcup_{i=\overline{1,N}:IA_i/FA=\varnothing} A_i.$$

We modify the set *A* according to the rule: $A=A/A^{(iter)}$.

Step 3. We find a project from the set $A^{(iter)}$, the duration of which is the smallest:

$$\Delta t_{\min} = \min_{i=\overline{1,N}: A_i \in A^{(iter)}} \Delta t_i.$$

Step 4. We change the current duration of project implementation $t_0=t_0+\Delta t_{\min}$ and for all $A_i \in A^{(iter)}$ we modify their duration as follows: $\Delta t_i = \Delta t_i - \Delta t_{\min}$.

Step 5. Tasks for which $\Delta t_i=0$ are marked completed as follows: $\forall i=\overline{1,N}: A_i \in A^{(iter)}$ if $\Delta t_i = 0$, then $FA=FA\cup\{A_i\}$, $A^{(iter)}=A^{(iter)}/\{A\}$.

Repeat Steps 3–5 until $A^{(iter)} \neq \varnothing$. Otherwise, proceed to Step 6.

Step 6. If $A\neq\varnothing$, then increase the iteration number by 1 ($iter=iter+1$) and proceed to Step 2. Otherwise, the algorithm is finished.

As a result of the iterative execution of the specified algorithm, the value $t_0$ will be equal to the minimum possible value of the duration of the project. Denote $T_{\min} = t_0$. Then the duration range of the project will be as follows: $T \in [T_{\min}; T_{n\ \max}]$.

The second stage is to determine the minimum possible cost of project implementation. A method of estimation of the minimum cost required to attract performers in the imple-mentation of a particular task is proposed. However, the issue of the cost of exploitation of material and technical resources, if necessary, can be solved in the same way.

For task $A_i$, function $\chi_i(s_q)$, is given that will allow calculating how many workers who can perform work $S_q$ must be involved at this stage. The function is constructed as follows:

$$\chi_i(s_q) = \begin{cases} \left[\dfrac{s_q}{\Delta t_i}\right], & \text{if } s_q = \left[s_q/\Delta t_i\right] \cdot \Delta t_i; \\ \left[\dfrac{s_q}{\Delta t_i}\right] + 1, & \text{otherwise}. \end{cases} \quad (7)$$

Let $X = (x_{jq})$, $j = \overline{1,M}$, $q = \overline{1,Q}$ be a Boolean matrix of the distribution of workers by work during the implementation of the task $A_i$. Moreover, $x_{jq} = 0$ if the employee $W_j$ will be involved to perform the work $S_q$ and $x_{jq} = 1$ otherwise.

Then, to find the minimum cost, it is necessary to solve the following boolean programming problem to find the components of vector $X$ in the following statement:

$$\sum_{j=1}^{M} \sum_{q=1}^{Q} x_{jq} c_{jq} \Delta t_i \to \min, \quad (8)$$

$$\sum_{q=1}^{Q} x_{jq} \leq 1, \forall j = \{1,2,...,M\}, \quad (9)$$

$$\sum_{j=1}^{m} x_{jq} = \chi_i(s_{iq}), \forall q \in \{1,2,...,Q\}, \quad (10)$$

$$x_{jq} \in \{0,1\} : \begin{cases} \text{if } x_{jq} = 1, \text{ then } w_{jq} = 1; \\ \text{if } w_{jq} = 0, \text{ then } x_{jq} = 0; \end{cases}$$

$$j = \overline{1,M}, q = \overline{1,Q}. \quad (11)$$

When solving problem (8) to (11), the following cases are possible:

– problem (8) to (11) has no solutions. This means that there are not enough existing employees to perform all the work of this stage. Then, the implementation of this task, and, consequently, the whole project is impossible. One possible solution in this case is to attract additional employees and re-solve problem (8) to (11);

– the problem has one solution $X^* = (x_{jq}^*)$. Then the minimum value of the cost of implementing the task $C_{i\min}$ can be calculated from (12):

$$C_{i\min} = \sum_{j=1}^{M} \sum_{q=1}^{Q} x_{jq}^* c_{jq} \Delta t_i; \quad (12)$$

– the problem has several solutions. This is possible if some workers are interchangeable. In this case, one of the solutions to problem (8) to (11) is chosen for the optimal solution and $C_{i\min}$ is calculated from (12).

After finding the minimum cost of each task, the minimum cost of implementing the entire project is calculated as follows:

$$\tilde{C}_{\min} = \sum_{i=1}^{N} C_{i\min}. \quad (13)$$

The «ideal point» for the task of building a hierarchy of tasks is a hierarchy that makes it possible to implement the entire project in time $T^* = \tilde{T}_{\min}$ with costs $C^* = \tilde{C}_{\min}$. In the case when the set $A$ has a topology of a «straight line», shown in Fig. 1, $b$, the achievement of the «ideal point» is possible by applying an iterative method of estimating the minimum value of the duration of the project, which also makes it possible to build a hierarchy. The cost of implementation will be determined from (13).

In other variants of the topologies of set $A$, the question of reaching or approaching the «ideal point» requires additional research.

An iterative method of sequential concessions for building a hierarchy of tasks is proposed. Given that at the previous stages the problem of determining the minimum cost of project implementation was solved, it can be argued that this project can be implemented in full by living employees.

The method consists in the consistent construction of a hierarchy of tasks, taking into account the available human resources.

Denote via $FA$ the set of tasks already completed; $DA$ – a set of tasks that are currently running; $WA_i$ – the set of employees involved in the task $A_i$ at the current time. For the first iteration, $T=0$, $FA=\varnothing$, $WA_i=\varnothing$, $DA=\varnothing$.

One iteration of the method for a given set of tasks $A$ and a set of workers $W$ can be described as follows:

Step 1. Using the method of estimating the minimum value of the project duration, we find $\tilde{T}_{\min}$.

Step 2. We fix the conditional start of the project $t_0=0$.

Step 3. From the elements of set $A$, we construct the set $A'$ including those elements that can already begin to be executed in this iteration, that is,

$$A' = \bigcup_{i=\overline{1,N}: IA_i/(FA \cup DA)=\varnothing} A_i.$$

We modify the set $A$ according to the rule: $A = A/A'$.

Step 4. For sets $A'$ and $W$ we solve the optimization problem:

$$\sum_{i: A_i \in A'} \left( \sum_{j: W_j \in W / \left( \bigcup_{i: A_i \in DA} WA_i \right)} \sum_{q=1}^{Q} y_{jq}^i c_{jq} \Delta t_i \right) \to \min, \quad (14)$$

$$\sum_{i: A_i \in A'} \sum_{q=1}^{Q} y_{jq}^i \leq 1, \forall j: W_j \in W / \left( \bigcup_{i: A_i \in DA} WA_i \right), \quad (15)$$

$$\sum_{j: W_j \in W / \left( \bigcup_{i: A_i \in DA} WA_i \right)} y_{jq}^i = \chi_i(s_{iq}),$$

$$\forall q \in \{1,2,...,Q\}, \forall i: A_i \in A', \quad (16)$$

$$y_{jq}^i \in \{0,1\} : \begin{cases} \text{if } y_{jq}^i = 1, \text{ then } w_{jq} = 1; \\ \text{if } w_{jq} = 0, \text{ then } y_{jq}^i = 0; \end{cases}$$

$$j: W_j \in W / \left( \bigcup_{i: A_i \in DA} WA_i \right), q = \overline{1,Q}, \quad (17)$$

where $y_{jq}^i = 1$, if worker $W_j$ performs work $S_q$ in task $A_i$ and $y_{jq}^i = 0$ otherwise.

When solving problem (14) to (17), the following cases are possible:

Case 1. The problem has no solution. This case occurs when there are not enough available employees to implement all tasks from $A'$. Possible management decisions are:

Solution 1.1. Attracting additional employees, which will make it possible adhere to the project implementation schedule but increase the cost of its implementation. If this decision is made, we add new employees to the set $W$ and return to Step 4.

Solution 1.2. Moving some tasks from set $A'$ to set $A$ and returning to Step 4. Making this decision will lead to an increase in the duration of the project. The magnitude of such an increase $\Delta \tilde{t} \geq \min\limits_{i:A_i \in A'} \Delta t_i$, where $A'$ is an updated set of tasks.

Case 2. The problem has a single solution $y_{jq}^{*i}$. The costs of implementing such a distribution of employees between tasks will be calculated from (18):

$$C^* = \sum_{i:A_i \in A'}\left(\sum_{j:W_j \in W/\left(\bigcup\limits_{i:A_i \in DA} WA_i\right)} \sum_{q=1}^{Q} y_{jq}^{*i} c_{jq} \Delta t_i\right). \quad (18)$$

To compare the result obtained with the minimum cost for each task $A_i \in A'$, it is necessary to solve problem (8) to (11). Then, the amount of monetary loss will be as follows: $\Delta \tilde{c} \geq C^* - \sum\limits_{i:A_i \in A'} C_{i\min}$. In this case, the following solutions are possible:

Solution 2.1. Monetary losses $\Delta \tilde{c}$ are permissible. Then we calculate $C = C + \Delta \tilde{c}$ and perform conversion (19) and proceed to Step 5.

$$WA_i = \bigcup_{\substack{j:W_j \in W/\left(\bigcup\limits_{\xi:A_\xi \in DA} WA_\xi\right),\\ \exists q \in \{1,2,\ldots,Q\}: y_{jq}^{*i}=1}} \{W_j\};$$

$$DA = DA \cup A'; \, A' = \emptyset. \quad (19)$$

Solution 2.2. In the event that monetary losses $\Delta \tilde{c}$ are not admissible, a decision similar to Solution 1.2 shall be implemented.

Step 5. We find a project from the $DA$ set, the duration of which is the smallest: $\Delta t_{\min} = \min\limits_{i=\overline{1,N}:A_i \in DA} \Delta t_i$.

Step 4. We change the current duration of the project $T = T + \Delta t_{\min}$ and for all $A_i \in DA$ modify their duration as follows: $\Delta t_i = \Delta t_i - \Delta t_{\min}$.

Step 5. Tasks for which $\Delta t_i = 0$ are marked completed as follows: $\forall i = \overline{1,N}: A_i \in DA$ if $\Delta t_i = 0$, then $FA = FA \cup \{A_i\}$, $W = W \cup WA_i$, $DA = DA/\{A_i\}$.

If $DA = \emptyset$, and $A = \emptyset$, then the algorithm is completed. Otherwise, proceed to the next iteration.

As a result of the execution of this algorithm, we obtain a hierarchy of tasks, the sequential implementation of which will make it possible to implement the project in time $T$ with costs $C$.

## 6. Discussion of results of the construction of methods for solving problems that arise at the project planning stage

The project planning stage is decisive in project activities [1]. Limited labor and financial resources, time limits for project implementation, and other additional conditions make this process difficult. The peculiarity of the initial conditions in this case, as a rule, is that the simultaneous observance of financial and time restrictions with the available labor resources is impossible, that is, the set of permissible solutions is empty. Also, decision-making is complicated when the set of optimal Pareto solutions has great power. As shown in [26], for the successful implementation of projects, it is necessary to apply the theory of related disciplines at the stages of their planning. This was done in our study.

Our mathematical model of the problem of constructing a hierarchy of tasks (1) to (5) makes it possible to formalize the following main components of the project planning process:
– tasks and interdependencies between them in the form of a tuple (2);
– performers, their competencies, and the cost of performing their work in the form (3);
– the amount of resources required for project implementation, financial costs, time limits, etc. in the form (1).

The task of constructing a hierarchy of tasks is to specify such a sequence (4) that would satisfy condition (5), as well as additional conditions provided by the DM. It is clear that there are cases when for a given set $A$ the hierarchy of tasks (4) can be constructed in several ways. This primarily depends on the topology of this set.

It is determined that the criteria for efficiency of the constructed hierarchy of tasks, in addition to compliance with condition (5), can be time, cost, and cost-time indicators of project implementation. To determine these indicators, the task of estimating the minimum value of the duration of the project implementation and the task of estimating the maximum required cost of the project implementation process were built. The DM selects performance criteria depending on the desired project implementation consequences and available resources.

Our iterative method for estimating the minimum value of the duration of the project makes it possible at the initial stages, when planning, to estimate the range of time over which the project can be implemented. It is clear that the minimum cost of project implementation can be achieved when all tasks are performed sequentially. This will make it possible to attract performers with a lower cost of work. The cost structure for the implementation of the project, given by (1) to (3), consists of the costs of operating material and technical resources and paying for the time of attracting performers. The resulting score can be chosen as a benchmark or «ideal point» in assessing the time efficiency of the project implementation process; it takes the value to which the method of building a hierarchy of tasks will strive.

Similar to time efficiency, a method for estimating the minimum project cost has been developed to be able to apply cost efficiency. According to the method, the problem of Boolean programming (8) to (11) is constructed. The method involves building a hierarchy of tasks without taking into account the restrictions on available human and logistical resources. Considered, according to this method, are only restrictions on the sequence of tasks.

In the course of solving this problem, it is possible to identify the lack of existing employees, which is a prerequisite for reviewing the input data and updating the process of solving these tasks. Consequently, this method makes it possible both to estimate the minimum cost of project implementation and to identify the fact of emptiness of the set of admissible solutions at the initial stages.

An iterative method of sequential concessions for building a hierarchy of tasks in a dialog mode, by sequentially solving optimization problems (14) to (17), has been built. If it is impossible to reach the «ideal point», unlike conventional optimization methods [14–16], the method makes it possible for DM to gradually move away from it, easing the requirements for the duration and cost of project implementation. In comparison with the method of successive concessions [24, 25], our method makes it possible to accept concessions not only to the components of the «ideal point» but also to change the initial data, in particular in the set of workers. The proposed solutions for this stage enable DM to act in accordance with different management strategies, which makes the proposed approach more universal and corresponds to the hypothesis of improving the quality of planning by establishing communication between responsible persons [27].

In the extreme case, a situation is possible when the hierarchy of tasks will not be built. This may occur when the available labor resources are not enough to implement the project in an acceptable time with a given budget.

The numerical experiments conducted on model examples allowed us to draw the following conclusions:

– in the case when the «ideal point» is achievable, the iterative method of constructing a hierarchy of tasks for a finite number of steps makes it possible to find the corresponding ordering of set $A$ with the distribution of tasks among existing performers;

– if the initial set of valid solutions is empty, depending on the initial data, if it is necessary to comply with time limits for project implementation, the cost of implementation increases. In some cases, by several times relative to the original;

– if there are not enough existing performers, then a conclusion on this will be obtained at the stage of solving problem (8) to (11);

– the number of iterations required to build a hierarchy of tasks, in general, depends on both the initial conditions and the magnitude of the concessions that the DM will make at each iteration.

Our models and methods could be effectively applied when planning projects in various spheres of human activity at the stage of distribution of work between their executors and establishing the sequence of implementation of individual tasks. Such projects should first be divided into separate tasks, indicating the links between them. To ensure that the hierarchy of tasks can be obtained, it is necessary that the minimum necessary financial resources and qualified task performers are potentially available for a given project.

At the next stages of the study, it is worth considering the models and conditions for the composition of several tasks into one in order to simplify the process of solving the problem of building a hierarchy of tasks.

## 7. Conclusions

1. Verbal and mathematical models have been constructed and tasks that arise at the stage of project planning in the context of distribution of performers between tasks were formalized. Among them are the task of building a hierarchy of tasks, the task of estimating the minimum values of the duration and cost of project implementation. The statements of these problems make it possible to take into account the relationships between the tasks, the implementation of which is necessary for the implementation of the entire project; competence and number of performers; duration of implementation of tasks; cost of work, etc. This makes it possible for DM in an understandable, formalized form to impose constraints that the solution of problems must meet.

In the future, similar to our models, it is possible to impose restrictions on the amount of material and technical resources, and so on.

2. A method for estimating the minimum duration of the project has been devised. The method is iterative, making it possible to take into account the relationships between different stages of project implementation.

A method for estimating the minimum cost of project implementation has been developed, which involves constructing a boolean programming problem to find such a distribution of work among performers that would minimize costs. The advantage of this method is that it makes it possible at the initial stages to identify the lack of performers of tasks and devise appropriate management decisions in this regard.

The application of the above methods makes it possible to calculate the components of the «ideal point» to which our algorithm of the iterative method for constructing a hierarchy of tasks will approach. For the case when the «ideal point» is unattainable, variants of solutions for the introduction of DM concessions on the mentioned values have been developed. Using this method, for a finite number of iterations, it is possible either to build a hierarchy of tasks in accordance with the provided restrictions on the duration and cost of project implementation, or to determine that this project cannot be implemented with given constraints.

## Conflicts of interest

The authors declare that they have no conflicts of interest in relation to the current study, including financial, personal, authorship, or any other, that could affect the study and the results reported in this paper.

## Funding

The study was conducted without financial support.

## Data availability

All data are available in the main text of the manuscript.